   \documentclass[12pt]{article}
   \usepackage{latexsym}
\else\oddsidemargin25mm\evensidemargin25mm\marginparwidth25mm\fi
\def\theequation{\arabic{section}.\arabic{equation}}

\renewcommand{\theequation}{\thesection.\arabic{equation}}
\begin{document}
\makeatletter \@addtoreset{equation}{section} \makeatother
\renewcommand{\theequation}{\thesection.\arabic{equation}}
\newcommand{\ft}[2]{{\textstyle\frac{#1}{#2}}}
\newcommand{\QED}{{\hspace*{\fill}\rule{2mm}{2mm}\linebreak}}
\def\dop{{\rm d}\hskip -1pt}
\def\bfone{\relax{\rm 1\kern-.35em 1}}
\def\bfzero{\relax{\rm I\kern-.18em 0}}
\def\inbar{\vrule height1.5ex width.4pt depth0pt}
\def\IC{\relax\,\hbox{$\inbar\kern-.3em{\rm C}$}}
\def\ID{\relax{\rm I\kern-.18em D}}
\def\IF{\relax{\rm I\kern-.18em F}}
\def\IK{\relax{\rm I\kern-.18em K}}
\def\IH{\relax{\rm I\kern-.18em H}}
\def\II{\relax{\rm I\kern-.17em I}}
\def\IN{\relax{\rm I\kern-.18em N}}
\def\IP{\relax{\rm I\kern-.18em P}}
\def\IQ{\relax\,\hbox{$\inbar\kern-.3em{\rm Q}$}}
\def\IR{\relax{\rm I\kern-.18em R}}
\def\IG{\relax\,\hbox{$\inbar\kern-.3em{\rm G}$}}
\font\cmss=cmss10 \font\cmsss=cmss10 at 7pt
\def\ZZ{\relax\ifmmode\mathchoice
{\hbox{\cmss Z\kern-.4em Z}}{\hbox{\cmss Z\kern-.4em Z}}
{\lower.9pt\hbox{\cmsss Z\kern-.4em Z}} {\lower1.2pt\hbox{\cmsss
Z\kern .4em Z}}\else{\cmss Z\kern-.4em Z}\fi}
\def\a{\alpha}
\def\b{\beta}
\def\d{\delta}
\def\e{\epsilon}
\def\g{\gamma}
\def\G{\Gamma}
\def\l{\lambda}
\def\L{\Lambda }
\def\s{\sigma}
\def\S{\Sigma}
\def\im {{\rm Im}\cN}
\def\cA{{\cal A}}
\def\cB{{\cal B}}
\def\cC{{\cal C}}
\def\cD{{\cal D}}
    \def\cF{{\cal F}}
    \def\cG{{\cal G}}
\def\cH{{\cal H}}
\def\cI{{\cal I}}
\def\cJ{{\cal J}}
\def\cK{{\cal K}}
\def\cL{{\cal L}}
\def\cM{{\cal M}}
\def\cN{{\cal N}}
\def\cO{{\cal O}}
\def\cP{{\cal P}}
\def\cQ{{\cal Q}}
\def\cR{{\cal R}}
\def\cV{{\cal V}}
\def\cW{{\cal W}}
%
%
%
\def\crr{\crcr\noalign{\vskip {8.3333pt}}}
\def\tilde{\widetilde}
\def\bar{\overline}
\def\us#1{\underline{#1}}
\let\shat=\hat
\def\hat{\widehat}
\def\hyp{\vrule height 2.3pt width 2.5pt depth -1.5pt}
\def\square{\mbox{.08}{.08}}
\def\Coeff#1#2{{#1\over #2}}
\def\Coe#1.#2.{{#1\over #2}}
\def\coeff#1#2{\relax{\textstyle {#1 \over #2}}\displaystyle}
\def\coe#1.#2.{\relax{\textstyle {#1 \over #2}}\displaystyle}
\def\half{{1 \over 2}}
\def\shalf{\relax{\textstyle {1 \over 2}}\displaystyle}
\def\dag#1{#1\!\!\!/\,\,\,}
\def\to{\rightarrow}
\def\notin{\hbox{{$\in$}\kern-.51em\hbox{/}}}
\def\shdot{\!\cdot\!}
\def\ket#1{\,\big|\,#1\,\big>\,}
\def\bra#1{\,\big<\,#1\,\big|\,}
\def\equaltop#1{\mathrel{\mathop=^{#1}}}
\def\Trbel#1{\mathop{{\rm Tr}}_{#1}}
\def\inserteq#1{\noalign{\vskip-.2truecm\hbox{#1\hfil}
\vskip-.2cm}}
\def\attac#1{\Bigl\vert
{\phantom{X}\atop{{\rm\scriptstyle #1}}\phantom{X}}}
\def\exx#1{e^{{\displaystyle #1}}}
\def\del{\partial}
\def\delbar{\bar\partial}
\def\nex#1{$N\!=\!#1$}
\def\dex#1{$d\!=\!#1$}
\def\cex#1{$c\!=\!#1$}
\def\eg{{\it e.g.}} \def\ie{{\it i.e.}}
%
\def\IE{\relax{{\rm I\kern-.18em E}}}
\def\cE{{\cal E}}
\def\cU{{\cal U}}
\def\rt{{\cR^{(3)}}}
\def\IGam{\relax{{\rm I}\kern-.18em \G }}
\def\IGa{\IA}
\def\cV{{\cal V}}
\def\Rt{{\cal R}^{(3)}}
\def\W{{\cal W}}
\def\tft#1{\langle\langle\,#1\,\rangle\rangle}
\def\IA{\relax{\hbox{{\rm A}\kern-.82em {\rm A}}}}
\let\picfuc=\fp
\def\hata{{\shat\a}}
\def\hatb{{\shat\b}}
\def\hatA{{\shat A}}
\def\hatB{{\shat B}}
\def\bv{{\bf V}}
\def\Fb{\overline{F}}
\def\nablab{\overline{\nabla}}
\def\Ub{\overline{U}}
\def\Db{\overline{D}}
\def\zb{\overline{z}}
\def\eb{\overline{e}}
\def\fb{\overline{f}}
\def\tb{\overline{t}}
\def\Xb{\overline{X}}
\def\Vb{\overline{V}}
\def\Cb{\overline{C}}
\def\Sb{\overline{S}}
\def\delb{\overline{\del}}
\def\Gammab{\overline{\Gamma}}
\def\Ab{\overline{A}}
\def\Anh{A^{\rm nh}}
\def\alphab{\bar{\alpha}}
\def\cy{Calabi--Yau}
\def\cabg{C_{\alpha\beta\gamma}}
\def\B{\S }
\def\Bh{\hat \S }
\def\Kh{\hat{K}}
\def\Knh{{\cal K}}
\def\A{\L }
\def\Ah{\hat \L }
\def\R{\hat{R}}
\def\V{{V}}
\def\T{T}
\def\Gammah{\hat{\Gamma}}
\def\twot{$(2,2)$}
\def\K{K\"ahler}
\def\rat{({\theta_2 \over \theta_1})}
\def\lv{{\bf \omega}}
\def\w{w}
\def\CP{C\!P}
\def\o#1#2{{{#1}\over{#2}}}
\newcommand{\be}{\begin{equation}}
\newcommand{\ee}{\end{equation}}
\newcommand{\ba}{\begin{eqnarray}}
\newcommand{\ea}{\end{eqnarray}}
\newtheorem{definizione}{Definition}[section]
\newcommand{\bd}{\begin{definizione}}
\newcommand{\ed}{\end{definizione}}
\newtheorem{teorema}{Theorem}[section]
\newcommand{\bth}{\begin{teorema}}
\newcommand{\eth}{\end{teorema}}
\newtheorem{lemma}{Lemma}[section]
\newcommand{\blem}{\begin{lemma}}
\newcommand{\elem}{\end{lemma}}
\newcommand{\brr}{\begin{array}}
\newcommand{\err}{\end{array}}
\newcommand{\nn}{\nonumber}
\newtheorem{corollario}{Corollary}[section]
\newcommand{\bcorol}{\begin{corollario}}
\newcommand{\ecorol}{\end{corollario}}
\def\twomat#1#2#3#4{\left(\begin{array}{cc}
 {#1}&{#2}\\ {#3}&{#4}\\
\end{array}
\right)}
\def\twovec#1#2{\left(\begin{array}{c}
{#1}\\ {#2}\\
\end{array}
\right)}
\begin{titlepage}
\hskip 5.5cm \hskip 1.5cm
\vbox{\hbox{CERN-TH/2001-376}\hbox{hep-th/0112192}\hbox{December,
2001}} \vfill \vskip 3cm
\begin{center}
{\Large {Consistent reduction of $N=2 \to N=1$ \\
 four dimensional supergravity coupled to matter}}\\
\vskip 1.5cm
  {\bf Laura Andrianopoli$^1$, Riccardo D'Auria$^1$ and Sergio Ferrara$^{1,2,3,4}$} \\
\vskip 0.5cm {\small $^1$ Dipartimento di Fisica, Politecnico di
Torino,\\
 Corso Duca degli Abruzzi 24, I-10129 Torino\\
and Istituto Nazionale di Fisica Nucleare (INFN) - Sezione di
Torino, Italy}\\
{\small $^2$ CERN Theoretical Division, CH 1211 Geneva 23,
Switzerland} \\
{\small $^3$ Istituto Nazionale di Fisica Nucleare (INFN) -
Laboratori Nazionali di Frascati}\\
{\small $^4$ Department of Physics
 \& Astronomy, University of California, Los Angeles, U.S.A.} \\
 \vspace{6pt}
\end{center}
\vskip 3cm \vfill
 {\bf Abstract}:
 {\small We analyze the constraints which follow both on the
 geometry and on the gauge sector for a consistent supergravity
 reduction of a general matter--coupled $N=2$ supergravity theory
 in four dimensions.
 These constraints can be derived in an elegant way by looking
 at the fermionic sector of the theory.
}
\end{titlepage}

In  this note we analyze the constraints which arise from a
consistent reduction of $N=2$ matter coupled supergravity
\cite{bw1},\cite{dlv},\cite{abcdffm}, with arbitrary gauging, to
$N=1$ standard matter coupled supergravity
\cite{cfgv},\cite{bagger}.
\par
This study may find applications in many physical situations, as
partial supersymmetry breaking \cite{bz1},\cite{aq}, brane
supersymmetry reduction \cite{bz2},\cite{bkvp}, string or M-
theories in presence of H-fluxes \cite{ps} -- \cite{tatar}.
\par
The main reason why a consistent reduction gives non trivial
constraints on the matter sector is due to the fact that the
second gravitino must be consistently eliminated from the
spectrum. This implies a condition:
\begin{equation}
\label{red}
  \delta_{\epsilon_1} \psi_{2\mu} =0|_{\epsilon_2 = 0}
\end{equation}
which must be integrable.
\par
In a general rigid supersymmetric theory the reduction $N=2 \to
N=1$ would give no constraint in the number of matter multiplets,
but only some restriction on their interactions. However this is
not the case in local supersymmetry because the second gravitino
multiplet generates some non linear couplings, which are required
to be absent if a consistent reduction may occur. \par A full
derivation of these constraints, also in higher $N$ theories, was
given recently \cite{lungo} by looking at the bosonic terms in the
local supersymmetry variations of the fermions.\par However, in
the full-fledged $N=2$ theory \cite{abcdffm}, the very same
variations contain three fermion terms:
\begin{equation}\label{3ferm}
  \delta f \sim ff\epsilon
\end{equation}
other that the bosonic terms
\begin{equation}\label{boson}
  \delta f \sim b \epsilon
\end{equation}
(here $f$ and $b$ denote generic fermion and boson fields).
\par
These terms, in the component formulation, have two different
origins. They come either from supercovariantization of bosonic
terms containing derivatives (such as all connection terms both of
space-time and of the scalar $\s$-model) or by elimination of
``auxiliary fields'' (or, in superspace, by solving Bianchi
identities of the gravitational multiplet coupled to the matter
multiplets.) \cite{abcdffm}.
\par
Note that these terms are crucial in the proof of local
supersymmetry of the lagrangian, since they induce five-fermion
terms in the supersymmetry variation of the lagrangian
\begin{equation}
  \delta_{\epsilon} {\mathcal L} = ffff\epsilon
\end{equation}
which must vanish by Fierz identities since they are purely
algebraic. The constraints on the three-fermion terms
(\ref{3ferm}) are much simpler to analyze since fermions have
simple transformation properties under the local symmetries of
the theory. By close inspection of these terms one can indeed
obtain a reduction of the matter sector which is precisely what
is obtained, by supersymmetry, from the constraints on the
reduced geometry analyzed in \cite{lungo}.
\par
We further note that the fermionic terms in the supersymmetry
variations do not depend on the gauging of the theory, so that the
restriction on the terms coming from the gauging must be still
analyzed separately and here we simply report their implications
on the reduced $N=1$ theory, as found in \cite{lungo}. However,
the consistency of the reduction in presence of gauging of any
isometry of the scalar manifold reflects, by supersymmetry, in the
occurrence of generalized Yukawa interactions, i.e. fermion
bilinear in the lagrangian. The consistent truncation requires
that all such terms, which are linear in the fermions which are
deleted, may not survive the reduction. This is seen to be a
consequence of the performed reduction.
\par
In this note we will analyze the reduction of the fermionic terms
in the supersymmetry variations and in the lagrangian, showing
that the results obtained are in perfect agreement with those
found at the bosonic level in \cite{lungo}.
\section{Fermionic contributions}
The supersymmetry reduction $N=2\to N=1$ is obtained by truncating
the $N=1$ spin $3/2$ multiplet containing the second gravitino
$\psi_{\mu 2}$ and the graviphoton.
\par
Here and in the following we use the notations both for $N=2$ and
$N=1$ supergravity as given in reference \cite{df}, the only
differences being that we use here world indices ${\mathcal
I},{\bar{\mathcal I}} =1,\cdots ,n_V$ and boldfaced gauge indices
${\bf \L }= 0,1,\cdots ,n_V$ for quantities in the $N=2$ vector
multiplets since we want to reserve the notation $\L$ and
 $i,\bar \imath$ for the indices of the reduced $N=1$ theory
 (see reference \cite{lungo}).
\par
 Let us write down the complete supersymmetry
transformation laws of the $N=2$ theory, including 3-fermions
terms \cite{abcdffm}:
 \vskip 0.5cm
 {\bf Supergravity transformation rules of
the (left--handed) Fermi  fields}:
\begin{eqnarray}
\delta\,\psi _{A \mu} &=& {\hat{\nabla}}_{\mu}\,\epsilon _A\,
 + \left [ {\rm i} \, g \,S_{AB}\eta _{\mu \nu}+
\varepsilon_{AB} \left(T^-_{\mu \nu} + U^+_{\mu \nu}
\right)\right] \gamma^{\nu}\epsilon^B +
\nonumber\\
&+& \left (A_{\mu A }^{\phantom{\mu A}B}+\g_{\mu\nu}A_{A
}^{\prime\nu\phantom{A}B} \right)\epsilon_B - \frac{1}4\left
(\partial_{{\mathcal J}} \cK \bar\l^{{\mathcal J}B}\epsilon_B -
h.c.
\right)\psi_{\mu A} +\nonumber\\
&-&\omega_{A |u}^{\phantom{A |u}B}\left ({\mathcal U}^{\alpha C
|u}\bar\zeta_\alpha \epsilon_C +{\mathcal U}^u_{\alpha C
}\bar\zeta^\alpha \epsilon^C \right)\psi_{\mu B}
 \label{gravtrasf} \\
\delta \,\lambda^{{\mathcal I}A}&=&
 {\rm i}\,\nabla _ {\mu}\, z^{{\mathcal I}}
\gamma^{\mu} \epsilon^A +G^{-{\mathcal I}}_{\mu \nu} \gamma^{\mu
\nu} \epsilon _B \varepsilon^{AB}\,+\, gW^{{\mathcal I}AB}\epsilon
_B
+\nonumber\\
&+& \frac{1}4\left (\partial_{{\mathcal J}} \cK \bar\l^{{\mathcal
J}B}\epsilon_B - h.c. \right)\lambda^{{\mathcal I}A}-\omega^A_{\ B
|u}\left ({\mathcal U}^{\alpha C |u}\bar\zeta_\alpha \epsilon_C
+{\mathcal U}^u_{\alpha C }\bar\zeta^\alpha \epsilon^C
\right)\lambda^{{\mathcal I}B}
+\nonumber\\
&-&\G^{{\mathcal I}}_{\ {\mathcal J}{\mathcal
K}}\bar\l^{{\mathcal K} B}\epsilon_B \lambda^{{\mathcal J}A}- {\rm
i}\lambda^{{\mathcal I}B}\psi_{\mu B}\g^\mu\epsilon^A + \frac{\rm
i}2 g^{{\mathcal I}\bar{\mathcal J}}\varepsilon^{AB}
C_{\bar{\mathcal J}\bar{\mathcal K}\bar{\mathcal L}}\bar
\l^{\bar{\mathcal K}}_B \l^{\bar{\mathcal
L}}_C\epsilon_D\varepsilon^{CD}
\label{gaugintrasf}\\
 \delta\,\zeta _{\alpha}&=&{\rm i}\,
{\mathcal U}^{B \beta}_{u}\, \nabla _{\mu}\,q^u \,\gamma^{\mu}
\epsilon^A \varepsilon _{AB}\,C_{\alpha  \beta} \,+\,g
N_{\alpha}^A\,\epsilon _A  +\nonumber\\
&+&\frac{1}4 \left (\partial_{{\mathcal J}} \cK \bar\l^{{\mathcal
J}B}\epsilon_B - h.c. \right)\zeta _{\alpha} - \Delta_{\alpha
|u}^{\phantom{\alpha |u}\beta}\left ({\mathcal U}^{\g A
|u}\bar\zeta_\g \epsilon_A +{\mathcal U}^u_{\g A }\bar\zeta^\g
\epsilon^A \right) \zeta_\beta + \nonumber\\
&-& {\rm i} \bar\zeta_\a \psi_{\mu A}\g^\mu \epsilon^A
\label{iperintrasf}
\end{eqnarray}
\vskip 0.5cm {\bf Supergravity transformation rules of the Bose
fields}:
\begin{eqnarray}
\delta\,V^a_{\mu}&=& -{\rm i}\,\bar {\psi}_{A
\mu}\,\gamma^a\,\epsilon^A -{\rm i}\,\bar {\psi}^A _
\mu\,\gamma^a\,\epsilon_A\\
\delta \,A^{{\bf \L } } _{\mu}&=& 2 \bar L^{{\bf \L } } \bar \psi
_{A\mu} \epsilon _B \epsilon^{AB}\,+\,2L^{{\bf \L }
}\bar\psi^A_{\mu}\epsilon^B \epsilon
_{AB}\nonumber\\
&+&{\rm i} \,f^{{\bf \L } }_{{\mathcal I} }\,\bar
{\lambda}^{{\mathcal I}A} \gamma _{\mu} \epsilon^B \,\epsilon
_{AB} +{\rm i} \, {\bar f}^{{\bf \L } }_{\bar{\mathcal I}}
\,\bar\lambda^{\bar{\mathcal I}}_A
\gamma _{\mu} \epsilon_B \,\epsilon^{AB} \label{gaugtrasf}\\
\delta\,z^{{\mathcal I}} &=& \bar{\lambda}^{{\mathcal I}A}\epsilon _A \label{ztrasf}\\
\delta\,z^{\bar{\mathcal I}}&=& \bar{\lambda}^{\bar{\mathcal I}}_A
\epsilon^A
\label{ztrasfb}\\
  \delta\,q^u &=& {\mathcal U}^u_{\alpha A} \left(\bar {\zeta}^{\alpha}
  \epsilon^A + C^{\alpha  \beta}\epsilon^{AB}\bar {\zeta}_{\beta}
  \epsilon _B \right). \label{quatertrasf}
 \end{eqnarray}
We have defined:
\begin{equation}
\hat{\nabla}_{\mu}\,\epsilon _A = {\mathcal D}_\mu \epsilon_A
 + \hat{\omega}_{\mu |
A}^{\phantom{\mu |A}B} \epsilon_B +\frac{\rm i}2 \hat{\cal Q}_\mu
\epsilon_A
\end{equation}
where ${\mathcal D}$ denotes the Lorentz covariant derivative (on
the spinors, ${\mathcal D}_\mu = \partial_\mu -\frac 14
\omega_\mu^{ab}\g_{ab}$),  and the $SU(2)$ and $U(1)$ 1-form
``gauged'' connections are respectively given by:
\begin{eqnarray}
 \hat{\omega}_{A}^{\phantom{A}B}&=&{\omega}_{A}^{\phantom{A}B} +
g_{({\bf \L })} \, A^{{\bf \L }} \,P^x_{{\bf \L }}\,
(\s^x)_{A}^{\phantom{A}B}\,,\label{oab}\\
\hat{\cal Q} &=& {\cal Q} + g_{({\bf \L })}\, A^{{\bf \L }}
\,P^0_{{\bf \L }} \,,\label{qu}\\
{\cal Q} &=& -\frac{\rm i}2 \left(\partial_{{\mathcal I}} \cK d
z^{{\mathcal I}} - \partial_{\bar{\mathcal I}} \cK d \bar
z^{\bar{\mathcal I}}\right)\label{qu2}
\end{eqnarray}
 ${\omega}_{A}^{\phantom{A}B}$, ${\cal Q}$ are the
 $SU(2)$ and $U(1)$ composite connections of the ungauged
theory. Moreover we have:
\begin{eqnarray}
\nabla_\mu z^{{\mathcal I}} &=& \partial_\mu z^{{\mathcal I}} +
g_{({\bf \L }
)} A^{{\bf \L }}_\mu k^{{\mathcal I}}_{{\bf \L } }\\
\nabla_\mu q^{u} &=& \partial_\mu q^{u} + g_{({\bf \L } )} A^{{\bf
\L }}_\mu k^{u}_{{\bf \L } }
\end{eqnarray}
where $k^{{\mathcal I}}_{{\bf \L } }$ and $k^u_{{\bf \L } }$ are
the Killing vectors of the $N=2$ special-K\"ahler manifold
$\cM^{SK}$ and of the quaternionic manifold $\cM^Q$ respectively.
They are related to the respective prepotentials by:
\begin{eqnarray}
k^{{\mathcal I}}_{{\bf \L } }&=&{\rm i} g^{{\mathcal I}\bar{\mathcal J}}
\partial_{\bar{\mathcal J}}P^0_{{\bf \L } }\\
k^u_{{\bf \L } }&=&\frac{1}{6 \l^2} \Omega^{x |
vu}\nabla_{v}P^x_{{\bf \L } } \,; \quad \l =-1
\end{eqnarray}
where $\Omega^{x}_{uv}$ is the $SU(2)$-valued curvature of the
connection $\omega_A^{\phantom{A} B}$, $\l$ is the scale of the
quaternionic manifold which in our conventions is fixed to the
value $\l =-1$ by supersymmetry (see ref. \cite{abcdffm}). The
prepotential $P^0_{{\bf \L } }$ satisfies:
\begin{equation}
  P^0_{{\bf \L } } L^{{\bf \L } }\,=\, P^0_{{\bf \L } }{\bar L}^{{\bf \L } }\,=\,0
\end{equation}
where $L^{\bf \L}$, together with its magnetic counterpart $M_{\bf
\L} \equiv \cN_{\bf\L \bf\S} L^{\bf \L}$, is the symplectic
section of the $Sp(2n_V )$ flat bundle over $\cM^{SK}$ in terms
of which the special-K\"ahler geometry is defined. Note that we
use throughout the paper the definition $f^{\bf \L}_{\mathcal I} =
\nabla_{\mathcal I} L^{\bf \L} = \partial_{\mathcal I}L^{\bf\L} +
\half \partial_{\mathcal I} \cK L^{\bf \L}$.
 $T^-_{\mu\nu}$ appearing in the supersymmetry
transformation law of the $N=2$ left-handed gravitini is the
``dressed'' graviphoton defined as:
\begin{eqnarray}\label{gravif}
T^-_{\mu\nu} &\equiv & 2{\rm i} {\rm Im} {\cal N}_{{\bf \L } {\bf
\S } } L^{{\bf \S } }\Bigl[  F_{\mu\nu}^{{\bf \L }  -} +
\bigl(L^{\bf \L } \bar\psi_{A\mu}\psi_{B\nu}\epsilon^{AB} + \bar
L^{\bf \L }
\bar\psi^{A\mu}\psi^{B\nu}\epsilon_{AB}+\nonumber\\
&-& {\rm i} \bar f^{\bf \L }_{\bar\mathcal I}
\bar\l^{\bar{\mathcal I}}_A\g_{[\nu}\psi_{\mu
]B}\epsilon^{AB}\bigr)^- +
 \frac 18 \nabla_{\mathcal K} f^{\bf\L}_{\mathcal L}
\bar\l^{{\mathcal K}A}\g_{\mu\nu}\l^{{\mathcal L}B} \epsilon_{AB}
-\frac 14 L^{\bf \L} \IC^{\a\b}\bar\zeta_\a\g_{\mu\nu}\zeta_\b
 \Bigr]
\end{eqnarray}
while
\begin{eqnarray}\label{vectors}
G^{{\mathcal I}-}_{\mu\nu} &\equiv & - g^{{\mathcal
I}\bar{\mathcal J}} {\rm Im } {\cal N}_{ {\bf \L }{\bf \S }
}\,\bar f^{{\bf \S } }_{\bar{\mathcal J}}
  \Bigl[  F_{\mu\nu}^{{\bf \L }  -} +  \bigl(L^{\bf
\L } \bar\psi_{A\mu}\psi_{B\nu}\epsilon^{AB} - {\rm i}  f^{\bf \L
}_{\mathcal K}
\bar\l^{{\mathcal K}A}\g_{[\nu}\psi^B_{\mu ]}\epsilon_{AB}+\nonumber\\
&-& {\rm i} \bar f^{\bf \L }_{\bar\mathcal I}
\bar\l^{\bar{\mathcal I}}_A\g_{[\nu}\psi_{\mu
]B}\epsilon^{AB}\bigr)^- +
 \frac 18 \nabla_{\mathcal K} f^{\bf\L}_{\mathcal L}
\bar\l^{{\mathcal K}A}\g_{\mu\nu}\l^{{\mathcal L}B} \epsilon_{AB}
-\frac 14 L^{\bf \L} \IC^{\a\b}\bar\zeta_\a\g_{\mu\nu}\zeta_\b
 \Bigr]
\end{eqnarray}
are the ``dressed'' field strengths of the vectors inside the
vector multiplets (the ``minus'' apex means taking the self-dual
part.). The ``auxiliary fields'' $A_{\mu A}^{\phantom{\mu A}B}$
and $A_{\mu A}^{\prime\phantom{\mu A}B}$ are defined as:
\begin{eqnarray}
A_{\mu A}^{\phantom{\mu A}B}&=&- \frac{\rm i}4 g_{\bar{\mathcal
I}{\mathcal J}}\left(\bar\l^{\bar{\mathcal I}}_A\g^\mu
\l^{{\mathcal J}B}- \delta_A^B \bar\l^{\bar{\mathcal I}}_C\g^\mu
\l^{{\mathcal J}C}\right) \\
A_{\mu A}^{\prime\phantom{\mu A}B}&=&\frac{\rm i}4
g_{\bar{\mathcal I}{\mathcal J}}\left(\bar\l^{\bar{\mathcal
I}}_A\g^\mu \l^{{\mathcal J}B}- \half \delta_A^B
\bar\l^{\bar{\mathcal I}}_C\g^\mu \l^{{\mathcal J}C}\right)
-\frac{\rm i}4 \delta_A^B \bar \zeta_\a \g^\mu \zeta^\a .
\end{eqnarray}
 Moreover the fermionic shifts $S_{AB}$, $
W^{{\mathcal I}\,AB}$ and $ N^A_{\alpha}$ are given in terms of
the prepotentials and Killing vectors of the quaternionic geometry
as follows:
\begin{eqnarray}\label{trapsi2}
S_{AB}&=& {\rm i} \frac {1}{2} P_{AB\, {\bf \L } } \,
 L^{{\bf \L } }
 \equiv {\rm i} \frac {1}{2} P^x_{{\bf \L }} \sigma^x_{AB}L^{{\bf \L }} \\
\label{tralam}
 W^{{\mathcal I}\,AB}&=& {\rm {i}} P^{AB}_{
{\bf \L } }\,g^{{\mathcal I}\bar{\mathcal J}} f^{{\bf \L }
}_{\bar{\mathcal J}} + \epsilon^{AB} k^{{\mathcal I}}_{{\bf \L }
}{\overline L}^{{\bf \L } }
\\
N^A_{\alpha}&=& 2\,{\mathcal U}^A_{\alpha u} \,k^u_{{\bf \L } }
{\overline L}^{{\bf \L } } \label{traqua1}\\
N^{\alpha}_A&=& -2 \,{\mathcal U}_{A u}^{\alpha} \,k^u_{{\bf \L }
} L^{{\bf \L } } \label{traqua2}
\end{eqnarray}
\par
Since we are going to compare the $N=2$ reduced theory with the
standard $N=1$ supergravity, we also quote the supersymmetry
transformation laws of the latter theory \cite{bw2},\cite{cfgv}.
We have, up to 3-fermions terms:
 \\
{\bf $N=1$ transformation laws}
\begin{eqnarray}
\label{trapsi1}\delta \psi_{\bullet \mu} &=& {\cal D}_{\mu}
\epsilon_{\bullet}+ \frac{\rm i}2 \hat Q_\mu\epsilon_{\bullet}
+{\rm {i}} L(z, \bar z) \gamma_{\mu} \varepsilon^{\bullet} \\
\label{trachi1} \delta \chi^i &=& {\rm {i}}
\left(\partial_{\mu}z^i + g_{(\L )} A^\L_\mu k^i_\L \right)
\gamma^{\mu} \varepsilon_{\bullet}  + N^{i}\varepsilon_{\bullet}\\
\label{tralamb1}\delta \lambda^{\L }_{\bullet} &=& {\mathcal
{F}}_{\mu \nu}^{(-) \L } \gamma^{\mu \nu} \varepsilon_{\bullet }
+{\rm {i}} D^{\L } \varepsilon_{\bullet} \\
\label{traviel1}\delta V^a_{\mu} &=& -\rm {i}
\psi_{\bullet} \gamma_{\mu} \varepsilon^{\bullet} + h.c.\\
\label{travec1}\delta A^{\L }_{\mu} &=& \rm{i}\frac {1}{2} \bar
\lambda^{\L }_{\bullet} \gamma_{\mu}
\varepsilon^{\bullet} + h.c.\\
 \label{trasca1}\delta z^i  &=& \bar \chi^i
\varepsilon_{\bullet}
\end{eqnarray}
where $\hat \cQ$ is defined in a way analogous to the $N=2$
definition (\ref{qu}) and:
\begin{eqnarray}\label{n1def}
L(z,\bar z)&=& W(z)\,e^{\frac {1}{2} {{\mathcal {K}}_{(1)}(z, \bar z)}}\,,\quad \nabla_{\bar\imath}L =0 \\
\label{defn} N^i &=& 2\, g^{i\bar\jmath} \,\nabla_{\bar\jmath}\, \bar L \\
 \label{dlambda} D^{\L } &=& - 2 ({\rm {Im}} f_{\L
\S })^{-1} P_{\S }(z,\bar z)
\end{eqnarray}
and $W(z),{\mathcal {K}}_{(1)}(z, \bar z) ,P_{\S }(z,\bar z),
f_{\L\S}(z)$ are the superpotential,  K\"{a}hler potential,
Killing prepotential and  vector kinetic matrix respectively
\cite{cfgv}, \cite{bw2}, \cite{bagger}. Note that for the
gravitino and gaugino fields we have denoted by a lower (upper)
dot left-handed (right-handed) chirality. For the spinors of the
chiral multiplets $\chi$, instead, left-handed (right-handed)
chirality is encoded via an
 upper holomorphic (antiholomorphic) world index
 ($\chi^i, \chi^{\bar\imath}$).
 \par
 Finally, we recall the equations defining special geometry:
 \begin{eqnarray}
 D_i V &=& U_i\nonumber\\
D_i U_j &=& i C_{ijk} g^{k\bar k}\bar U_{\bar k}\nonumber\\
D_i U_{\bar\jmath} &=& g_{i \bar\jmath } \bar V\nonumber\\
D_{i} \bar V &=& 0 \label{geospec}
\end{eqnarray}
where
\begin{equation}V = (L^{\bf \Lambda},M_{\bf \Lambda})\ ,\quad  U_i =D_i V=(f_i^{\bf \Lambda},h_{{\bf\Lambda} i})\ \ \ \ \
{\bf \Lambda}=0,\ldots,n \label{sezio}; \end{equation}
\begin{equation}
M_{\bf\L} = \cN_{\bf\L\S} L^{\bf\S}\, , \quad h_{{\bf\Lambda}
i}=\bar \cN_{\bf\L\S}f_i^{\bf \Lambda}
\end{equation}
and $\cN_{\bf \L\S}$ is the kinetic vector matrix.
\vskip 5mm
 {\bf Gravitino reduction}
 \\
 To perform the truncation
we set $A$=1 and 2 successively, putting $\psi_{2\mu}
=\epsilon_2=0$, and we get from equation  (\ref{gravtrasf}):
\begin{eqnarray}
\label{reductio} \delta\,\psi _{1 \mu} &=& {\mathcal D}_\mu
\epsilon_1  -\hat{\cal Q}_\mu \epsilon_1 - \hat{\omega}_{\mu |
1}^{\phantom{\mu |A}1} \epsilon_1  \,
 +  {\rm i} \, g \,S_{11}\eta _{\mu \nu}
 \gamma^{\nu}\epsilon^1
+ \left (A_{\mu 1 }^{\phantom{\mu A}1}+\g_{\mu\nu}A_{1
}^{\prime\nu\phantom{A}1} \right)\epsilon^1 +\nonumber\\
&-&\omega_{1 |u}^{\phantom{A |u}1}\left ({\mathcal U}^{\alpha 1
|u}\bar\zeta_\alpha \epsilon_1 +{\mathcal U}^u_{\alpha 1
}\bar\zeta^\alpha \epsilon^1 \right)\psi_{\mu 1}- \frac{1}4\left
(\partial_{{\mathcal J}} K \bar\l^{{\mathcal J}1}\epsilon_1 -
h.c. \right)\psi_{\mu 1}
\end{eqnarray}
while, for consistency:
\begin{eqnarray}
\delta\,\psi _{2 \mu}\equiv 0 &=&  - \hat{\omega}_{\mu |
2}^{\phantom{\mu |A}1} \epsilon_1 +
  \left [ {\rm i} \, g \,S_{21}\eta _{\mu \nu}-\left(T^-_{\mu \nu} + U^+_{\mu \nu}
  \right)\right]
 \gamma^{\nu}\epsilon^1 +\nonumber\\
 &+& \left (A_{\mu 2 }^{\phantom{\mu A}1}+\g_{\mu\nu}A_{2
}^{\prime\nu\phantom{A}1} \right)\epsilon^1 -\omega_{2 |u}^{\phantom{A
|u}1}\left ({\mathcal U}^{\alpha 1 |u}\bar\zeta_\alpha \epsilon_1
+{\mathcal U}^u_{\alpha 1 }\bar\zeta^\alpha \epsilon^1 \right)
\psi_{\mu 1}
\end{eqnarray}
\par
Comparing (\ref{trapsi1}) with (\ref{reductio}), we learn that we
must identify:
\begin{eqnarray}
\psi_{1\mu} &\equiv & \psi_{\bullet \mu} \\
\epsilon_1&\equiv & \epsilon_{\bullet }.
\end{eqnarray}
 Furthermore, for a consistent truncation we must
set to zero all the following structures:
\begin{eqnarray}
&&T^-_{\mu \nu} = 0 \label{tmunu}\\
&&S_{21} =0 \label{s21}\\
&&\hat{\omega}_{\mu | 2}^{\phantom{\mu |A}1} =0\label{o21}\\
&&\hat{\omega}_{u | 2}^{\phantom{\mu |A}1}{\mathcal U}^u_{\alpha 1
} \bar\zeta^\alpha \epsilon^1 =0
 \label{f1}\\
 &&U^+_{\mu \nu} = - \frac {\rm i}4 \IC_{\a\b} \bar\zeta^\a
\g_{\mu\nu} \zeta^\b =0\label{f2}\\
 &&A_{\mu 2 }^{\phantom{\mu A}1} = -\frac {\rm i}4
 g_{\bar{\mathcal I} {\mathcal J}}\bar\l^{\bar{\mathcal
 I}}_2\g_\mu
 \l^{ {\mathcal J} 1}= - A_{\mu 2 }^{\prime\phantom{\mu A}1} =0 \label{f3}
\end{eqnarray}
We note that the expression ``equal to zero'' in (\ref{tmunu}) -
(\ref{f3}) has to be intended in a weak sense, as a condition to
be true on the reduced $N=1$ theory. \par
 Let us analyze in
particular the constraints (\ref{tmunu}), (\ref{f1}), (\ref{f2}),
(\ref{f3}) containing 3 fermions contributions.
\\
We first consider the implications of these constraints on the
hypermultiplet sector. Equations (\ref{f2}) and (\ref{tmunu})
impose to truncate out half of the hypermultiplets. Indeed, let us
decompose the symplectic index $\a \to ( I , \dot I )$, so that
we can write the symplectic matrix $\IC_{\a\b}$ as:
\begin{equation}
 \pmatrix{
   0 & \bfone_{I\dot I} \cr
   - \bfone_{\dot II} & 0}
\end{equation}
Then, equation (\ref{f2}) becomes:
\begin{equation}
 \delta_{I\dot I} \bar\zeta^I
\g_{\mu\nu} \zeta^{\dot I} =0
\label{f2'}
\end{equation}
which is an orthogonality condition between the set of
$\{\zeta^I\}$ and $\{\zeta^{\dot I}\}$. A particular solution is
to take $\zeta^I\neq 0$, and then:
\begin{equation}
\zeta^{\dot I} =0,\label{simple}
\end{equation}
that is at least half of the hypermultiplets have to be projected
out in the truncation. More generally, we could decompose the
indices as $I=(f,g)$; $\dot I = (\dot f,\dot g)$ (with $f, \dot f
= 1, \cdots , k\leq n_H$; $g,\dot g = 1,\cdots , n_H -k$) and, for
$\zeta^f \neq 0$; $\zeta ^{\dot g}\neq 0$ eq. (\ref{f2'}) gives
\begin{equation}
\zeta^g =0 \,\,;\quad \zeta^{\dot f} =0 \label{nonsimple}
\end{equation}
together with their scalar partners that, as we easily see when
looking at the hyperini transformation laws, are respectively:
\begin{equation}
{\mathcal U}^{1g }_u d q^u =0\,; \,\quad {\mathcal U}^{1\dot f }_u
d q^u =0\,; \quad \left({\mathcal U}^{1g} = ({\mathcal U}^{2 \dot
g })^* \,; \,\quad {\mathcal U}^{1\dot f } = ({\mathcal U}^{2 f
})^*\right).\label{partners}
\end{equation}
However, by a symplectic rotation we can always choose a basis
where $\dot g =0$, $f=I$. As we will show in the following when
looking at the hyperini transformation law reduction, there is no
loss of generality by adopting the simpler choice (\ref{simple})
(that is $\dot g =0$, $f=I$), as we will actually do in the
following. Therefore in the rest of the paper we will treat the
case $f=I$, where the only vielbein surviving on the submanifold
$\cM^{KH} \subset \cM^Q$ are:
\begin{equation}
{\mathcal U}^{1I } =({\mathcal U}^{2 \dot I })^* \label{surv}
\end{equation}
while:
\begin{equation}
{\mathcal U}^{2I } =({\mathcal U}^{1 \dot I })^*=0. \label{surv2}
\end{equation}
\par
 Now we can make a choice of coordinates on
the quaternionic manifold $q^u = (w^s, n^t)$, such that  the
$n^t$ are the coordinates truncated out, and set, as a basis of
vielbein for the submanifold spanned by the scalars of the
surviving hypermultiples:
\begin{eqnarray}
P_I &=& P_{Is} dw^s \equiv  \sqrt{2} {\mathcal U}_{I 1 u} dq^u \\
\bar P_{\dot I} &=& P_{\dot I\bar s} d\bar w^{\bar s} \equiv
\sqrt{2} {\mathcal U}_{\dot I 2 u} dq^u .
\end{eqnarray}
\\
With this position, equation (\ref{f1}) is now easy to interpret.
It can be rewritten as:
\begin{equation}
\hat{\omega}_{u | 2}^{\phantom{\mu |2 }1}{\mathcal U}^u_{I 1 }
\bar\zeta^I \epsilon^1 = \hat{\omega}_{\bar s | 2}^{\phantom{\mu
|A}1}{\mathcal U}^{\bar s}_{I 1 } \bar\zeta^I \epsilon^1  = 0
\end{equation}
which gives a condition on the component of the $SU(2)$
connection:
\begin{equation}
\hat{\omega}_{\bar s | 2}^{\phantom{\mu |A}1}=0. \label{f1'}
\end{equation}
This condition , obtained from the fermion-bilinear  equation
(\ref{f1}), coincides with the bosonic constraint (\ref{o21})
analyzed in reference \cite{lungo}. Indeed eq. (\ref{o21}) is
more properly written, in the appropriate basis, as:
\begin{equation}
\hat{\omega}_{ 2}^{\phantom{\mu
|A}1}|_{\cM^{KH}}=\hat{\omega}_{\bar s | 2}^{\phantom{\mu
|A}1}d\bar w^{\bar s} =0,
\end{equation}
which is satisfied by (\ref{f1'}).
 When looking at the explicit expression of
the field-strength of the $SU(2)$ connection (whose component
$\Omega_{ 2}^{\phantom{A}1}$ has to be zero for consistency):
\begin{equation}
\Omega_{ 2}^{\phantom{A}1} \, \equiv \, d \omega_{
2}^{\phantom{A}1} + \omega_{ 2}^{\phantom{A}A} \wedge
\omega_{A}^{\phantom{A}1} = \\ {\rm i}\, \lambda
 {\mathcal U}_{\alpha 2} \wedge {\mathcal
U}^{\a 1} = {\rm i}\, \lambda
 \left({\mathcal U}_{I 2} \wedge {\mathcal
U}^{I 1}+{\mathcal U}_{\dot I 2} \wedge {\mathcal U}^{\dot I
1}\right) =0 \label{curform1}
\end{equation}
we see that it is automatically satisfied by the position
(\ref{partners}).
\par
The surviving $U(1)$ curvature $\Omega_1^{\ 1} =\Omega^3\,
(\sigma^3)_1^{\ 1}$ is instead different from zero and defines
(one half) the K\"ahler 2-form on $\cM^{KH}$, so that we may
introduce complex coordinates $w^s$ and K\"ahler metric such that
$\Omega^3 = \frac{\rm i}2 g_{s\bar s}dw^s \wedge d\bar w^{\bar s}
$ \cite{lungo}.  This does not exhaust the restrictions on the
quaternionic manifold $\cM^Q$, since, as we will see in the
analysis of the fermionic sector of the hyperini transformation
laws, extra constraints on the symplectic part of the
quaternionic curvature have to be imposed.
\par
Let us now come to the reduction of the $N=2$ vector multiplets.
To understand condition (\ref{f3}), let us observe that the $n_V$
$N=2$ vector multiplets $( A_\mu , \l^{ A}, z )^{\mathcal I}$
(${\mathcal I} =1,\cdots n_V$) decompose to $N=1$ chiral
multiplets $( \l^{ 1}, z )^{\mathcal I}$  and $N=1$ vector
multiplets $( A_\mu, \l^{2} )^{\mathcal I}$ \cite{lungo}. Let us
suppose that in the reduction the number of chiral multiplets
coming from $N=2$ vector multiplets is $n_C \leq n_V$. We then
have to decompose the indices ${\mathcal I} \to (i,\a )$, where $i
=1,\cdots , n_C$ and $\a =n_C +1 , \cdots ,{\mathcal I}$, and the
chiral multiplets are labeled as $( \l^{ 1}, z
)^{i}$ (while $(\l^{1},z)^\a =0$).\\
 Then eq. (\ref{f3}) can be rewritten as:
\begin{equation}
g_{\bar{\imath}j}\bar\l^{\bar{\imath}}_2 \l^{ {j} 1}+
g_{\bar{\a}j}\bar\l^{\bar{\a}}_2 \l^{ {j} 1}= 0
\end{equation}
which is an orthogonality condition between the $N=1$ chiral and
vector multiplets coming from the $N=2$ vector multiplets,
satisfied for:
\begin{eqnarray}
\l^{i2} &=&0\\
g_{i\bar\a}&=&0.\label{mistametr}
\end{eqnarray}
The previous equations imply that if the $N=1$ chiral multiplets
have indices $i=1,\cdots n_C \leq n_V$, then the $N=1$ vector
multiplets take the  complementary indices $\a = \L =1,\cdots n'_V
=n_V- n_C$. As a consequence, the scalar partners of the chiral
fermions $\l^{i1}$ span a K\"ahler manifold $\cM_R \subset
\cM^{SK}$ of complex dimension $n_C$.
\par
Furthermore, the three fermion terms in eq. (\ref{tmunu}) and
(\ref{gravif}) containing $N=2$ gaugini  impose conditions on the
scalar sector of the theory. Indeed (\ref{tmunu}),
(\ref{gravif})  give:
\begin{eqnarray}
 {\rm Im} {\cal N}_{{\bf \L } {\bf \S } } L^{{\bf \S } }
  \bar f^{\bf \L }_{\bar\a} \bar\l^{\bar{\a}}_2\g_{[\nu}\psi_{\mu ]1} =0 &\Rightarrow
& {\rm Im} {\cal N}_{{\bf \L } {\bf \S } } L^{{\bf \S } }
  \bar f^{\bf \L }_{\bar\a} =0 \label{vec1}\\
{\rm Im} {\cal N}_{{\bf \L } {\bf \S } } L^{{\bf \S } }
 \nabla_{i} f^{\bf\L}_{\a}
\bar\l^{i 1}\g_{\mu\nu}\l^{{\a}2}  =0& \Rightarrow &{\rm Im}
{\cal N}_{{\bf \L } {\bf \S } } L^{{\bf \S } }
 \nabla_{i} f^{\bf\L}_{\a} =0 \label{vec2}
\end{eqnarray}
Equation (\ref{vec1}) is an orthogonality condition between the
set $\{ L^{\bf \L}\}$ and the set $\{f^{\bf \L}_\a \}$. By
decomposing the vector indices ${\bf \L}$ as ${\bf \L} \to (\L ,
X)$, (with $\L =1,\cdots ,  n'_V$, $X=0,1,\cdots  n_C =n_V-n_V'$)
it becomes:
\begin{equation}
{\rm Im} {\cal N}_{{ \L } { \S } } L^{{ \S } }
  \bar f^{\L }_{\bar\a} +{\rm Im} {\cal N}_{{ \L } {X } } L^{X }
  \bar f^{ \L }_{\bar\a} +{\rm Im} {\cal N}_{X { \L } } L^{{\L } }
  \bar f^{X }_{\bar\a} +{\rm Im} {\cal N}_{XY } L^{Y }
  \bar f^{X }_{\bar\a} =0 \label{vec1'}
  \end{equation}
  A consistent solution of eq. (\ref{vec1'}) is easily found by
  setting:
  \begin{eqnarray}
  L^\L &=& 0 \label{spec1}\\
  f^X_\a &=& 0 \label{spec2}\\
  {\rm Im}\cN_{\L X} &=& 0
  \label{spec3}
  \end{eqnarray}
We observe that, since $\cN_{\L X} $ is anti-holomorphic on
$\cM_R$, equation (\ref{spec3}) still allows a constant, purely
real, term $\cN_{\L X} =C_{\L X}$ that we do not discuss here.
\par
  With the same decomposition of indices the second equation
(\ref{vec2})  gives:
\begin{equation}
{\rm Im} {\cal N}_{{ \L } { \S } } L^{{ \S } }
  \nabla_i f^{\L }_{\a} +{\rm Im} {\cal N}_{{ \L } {X } } L^{X }
  \nabla_i  f^{ \L }_{\a} +{\rm Im} {\cal N}_{X { \L } } L^{{\L } }
  \nabla_i  f^{X }_{ \a} +{\rm Im} {\cal N}_{XY } L^{Y }
  \nabla_i  f^{X }_{ \a} =0 \label{vec2'}
  \end{equation}
  which is satisfied (in a way consistent with (\ref{spec1}) -
  (\ref{spec3})) with the further constraint:
  \begin{equation}
\nabla_i  f^{X }_{ \a}= {\rm i} C_{ij\a} f^X_{\bar\jmath}
g^{j\bar\jmath} = 0 \, \Rightarrow \, C_{ij\a} =0 \label{spec4}
\end{equation}
where we have used the special geometry relation (\ref{geospec})
defining $C_{\mathcal IJK}$.
 This solution tells us
that the reduced manifold $\cM_R$ is a special-K\"ahler manifold
with symplectic sections $(L^X, M_X)$. Indeed we have, recalling
the differential identities satisfied by the symplectic sections
of the $N=2$ parent theory, that the 3 equations (\ref{spec1}) -
(\ref{spec3}) induce on $\cM_R$ the special-K\"ahler structure
with indices ${\bf \L}$ restricted to $X$. Other possible
solutions to equations (\ref{spec1}) - (\ref{spec3}) are not
compatible with supersymmetry, as can be easily ascertained by
looking at the bosonic sector (see \cite{lungo}).
\par
 Given the conditions found above, let us now compute the reduction of the
complete transformation laws for the spin one half fermions.
\vskip 5mm
 {\bf Hypermultiplets reduction}\\
 The $N=2$
hyperini supersymmetry transformation law reduces to:
\begin{eqnarray}
\delta\,\zeta _{I}&=&{\rm i}\, {\mathcal U}^{2 \dot J}_{u}\,
\nabla _{\mu}\,q^u \,\gamma^{\mu} \epsilon^1 \,\delta_{I\dot J}
\,+\,g
N_{I}^1\,\epsilon _1  +\nonumber\\
&+&\frac{1}4 \left (\partial_{{\mathcal J}} K \bar\l^{{\mathcal
J}1}\epsilon_1 - h.c. \right)\zeta _{I} -
\Delta_{I|u}^{\phantom{I|u}J}\left ({\mathcal U}^{K 1
|u}\bar\zeta_K \epsilon_1 +{\mathcal U}^u_{K1 }\bar\zeta^K
\epsilon^1 \right) \zeta_J + \nonumber\\
&-& {\rm i} \bar\zeta_I \psi_{\mu 1}\g^\mu \epsilon^1
\label{iperchirtrasf}
\end{eqnarray}
while for consistency we have to impose:
\begin{eqnarray}
\delta\,\zeta _{\dot I}=0 &=&{\rm i}\, {\mathcal U}^{2 J}_{u}\,
\nabla _{\mu}\,q^u \,\gamma^{\mu} \epsilon^1 \,\delta_{\dot IJ}
\,+\,g
N_{\dot I}^1\,\epsilon _1  +\nonumber\\
&-& \Delta_{\dot I|u}^{\phantom{I|u}J}\left ({\mathcal U}^{K 1
|u}\bar\zeta_K \epsilon_1 +{\mathcal U}^u_{K1 }\bar\zeta^K
\epsilon^1 \right) \zeta_J + \nonumber\\
&+& {\rm i} \delta_{J\dot I }{\mathcal U}^{ J 2 }_u \nabla_\mu
q^u  \g^\mu \epsilon^1 \label{buttovia}
\end{eqnarray}
We find therefore the consistency conditions:
\begin{eqnarray}
\Delta_{\dot I}^{\phantom{I|u}J}&=&0\label{hyp1}\\
{\mathcal U}^{ J 2 }_u \nabla_\mu q^u&=&0\label{hyp2}\\
N^1_{\dot I}=0\label{hyp3}
\end{eqnarray}
Eq. (\ref{hyp1})  reduces the holonomy of the quaternionic scalar
manifold from $Sp(2n_H)$ to $U(n_H)$, a condition necessary for
the validity of the truncation, since the manifold has to reduce
to a K\"ahler-Hodge one.\\
We note that, if we had chosen the more general configuration
(\ref{nonsimple}), we had found instead of (\ref{hyp1}) the
holonomy constraints:
\begin{equation}
\Delta_{g}^{\ \ f} =0\, ; \quad \Delta_{g}^{\ \ \dot g } =0\, ;
\quad \Delta_{\dot f}^{\ \ \dot g} =0 \, ; \quad \Delta_{\dot
f}^{\ \ f} =0 \label{storti}
 \end{equation}
Working out the curvatures associated to these components of the
$Sp(2n_H)$ connection it is easy to see that they in fact
reconstruct the full curvatures of the group $U(n_H)$, embedded
however into $Sp(2n_H)$ in a different way from the standard one
related to the choice (\ref{surv}), (\ref{surv2}). In group
theoretical terms, if we set $f, \dot f =1,\cdots k$; $g, \dot g
= 1,\cdots n_H-k$, we find that the constraints (\ref{storti})
correspond to the decomposition:
\begin{equation}
Adj(Sp(2n_H)) \to Adj(U(k)) + Adj(U(n_H-k)) + 2(k,n_H
-k)\label{storto}
\end{equation}
Actually, in equation (\ref{storti}) we recognize that the r.h.s.
is in fact the adjoint of $U(n_H)$, which is however decomposed
with respect to its maximal subgroup $U(k)\times U(n_H -k)$. In
the sequel we refer only to the simpler choice (\ref{simple}).
\par
We stress the fact that the necessary condition (\ref{hyp1})
found above implies a further geometric constraint for the
consistency of the truncation. Indeed, as it has been analyzed in
\cite{lungo} by using the Frobenius theorem, in order for the
equations (\ref{surv2}) and (\ref{hyp1})to give a consistent
truncation, the quaternionic manifold cannot be generic; in
particular, the completely symmetric tensor
$\Omega_{\alpha\beta\gamma\delta}\in Sp(2n_H)$, appearing in the
$Sp(2n_H)$ curvature, must obey the following constraint:
\begin{equation}
 \Omega_{ \dot I\dot J K\dot L}\, =\, 0. \label{omega condition}
  \end{equation}
\par
 Eq. (\ref{hyp2}) is automatically satisfied with the choice of basis (\ref{hyp1}).
 Indeed it means that the scalar
partners of the $\zeta_{\dot I}$ have to be truncated out, since
they span the orthogonal complement to the retained submanifold:
\begin{equation}
{\mathcal U}^{ J 2 }_u \nabla_\mu q^u |_{\cM_{KH}}={\mathcal U}^{
J 2 }_t \nabla_\mu n^t|_{\cM_{KH}}=0 .
\end{equation}
We can now define chiral spinors with world indices:
\begin{equation}
\zeta^s \equiv  {\sqrt{2}} P^{I, s} \zeta_I
\end{equation}
 and we
find, for the tranformation law of the $\zeta^s$:
\begin{eqnarray}
\delta\,\zeta^{s}&=&{\rm i}\,  \nabla _{\mu}\,w^s \,\gamma^{\mu}
\epsilon^1  \,+\,g
N^s \,\epsilon _1  +\nonumber\\
&+&\frac{\rm i}4 \left (\partial_{{\mathcal J}} K
\bar\l^{{\mathcal J}1}\epsilon_1 - h.c. \right)\zeta^s
-\G^s_{s's''}\,\zeta^{s''}
\,\bar\zeta^{s'} \epsilon_1  + \nonumber\\
&-& {\rm i} \bar\zeta^s \psi_{\mu 1}\g^\mu \epsilon^1
\label{iperchirtrasf'}
\end{eqnarray}
with $N^s \equiv \sqrt{g}P^{I,s}N^1_I$ and $\G^s_{s's''}
=\delta_I^J P_{K ,s'} \partial_{s''} P^{K ,s}-  P^{I,s}
\Delta_{I|s'}^{\phantom{I|u}J} P_{J,s''}$.
\vskip 5mm
{\bf Vector multiplets reduction} \\
Let us now consider the
reduction of the gaugini transformation law. From eq.
(\ref{gaugintrasf}) we find:
\begin{eqnarray}
\delta \,\lambda^{i 1}&=&
 {\rm i}\,\nabla _ {\mu}\, z^{i}
\gamma^{\mu} \epsilon^1 \,+\, gW^{i11}\epsilon _1
+\nonumber\\
&+& \frac{1}4\left (\partial_{j} K \bar\l^{j 1}\epsilon_1 - h.c.
\right)\lambda^{i 1}-\left (\omega^1_{\ 1 |s}\bar\zeta^s
\epsilon_1 +\omega^1_{\ 1 |\bar s}\bar\zeta^{\bar s} \epsilon^1
\right)\lambda^{i1} -\G^{i}_{\ jk }\bar\l^{k 1}\epsilon_1
\lambda^{j 1}
+\nonumber\\
&-&{\rm i}\lambda^{i 1}\psi_{\mu 1}\g^\mu\epsilon^1 - \frac{\rm
i}2 g^{i\bar{j}}\,C_{\bar{j}\bar{\a}\bar{\b}}\bar \l^{\bar{\a}}_2
\l^{\bar{\b}}_2\epsilon_1 \label{gauchirtrasf}\\
\delta \,\lambda^{\a 2}&=&
 - G^{-\a}_{\mu \nu} \gamma^{\mu
\nu} \epsilon _1 \,+\, gW^{\a 21}\epsilon _1
+\nonumber\\
&+& \frac{1}4\left (\partial_{j} K \bar\l^{j1}\epsilon_1 - h.c.
\right)\lambda^{\a 2}-\left (\omega^2_{\ 2 |s}\bar\zeta^s
\epsilon_1 +\omega^2_{\ 2 |\bar s}\bar\zeta^{\bar s}\epsilon^1
\right)\lambda^{\a 2} -\G^{\a }_{\ \b i}\bar\l^{i 1}\epsilon_1
\lambda^{\b 2}
+\nonumber\\
&+& \frac{\rm i}2 g^{\a \bar{\b}}\,
C_{\bar{\b}\bar{i}\bar{\g}}\bar \l^{\bar{i}}_1
\l^{\bar{\g}}_2\epsilon_1 . \label{gaugautrasf}
\end{eqnarray}
while for consistency we have to impose:
\begin{eqnarray}
\delta \,\lambda^{{\a}1}&=& 0\,=\,
 {\rm i}\,\nabla _ {\mu}\, z^{\a}
\gamma^{\mu} \epsilon^1 \,+\, gW^{\a 11}\epsilon _1 -\G^{\a}_{\
ij}\bar\l^{j 1}\epsilon_1 \lambda^{i1}- \frac{\rm i}2
g^{\a\bar{\b}} C_{\bar{\b}\bar{\g}\bar{\delta}}\bar
\l^{\bar{\g}}_2 \l^{\bar{\delta}}_2\epsilon_1
\label{viagau1}\\
\delta \,\lambda^{i2}&=&0\,=\,
 - G^{-i}_{\mu \nu} \gamma^{\mu
\nu} \epsilon _1 \,+\, gW^{i21}\epsilon _1 -\G^{i}_{\ \a
j}\bar\l^{j1}\epsilon_1 \lambda^{\a 2} + \frac{\rm i}2
g^{i\bar{\jmath}} C_{\bar{\jmath}\bar{k}\bar{\a}}\bar
\l^{\bar{k}}_1 \l^{\bar{\a}}_2\epsilon_1\label{viagau2}
\end{eqnarray}
that implies that, on the reduced theory, the following quantities
have to be zero:
\begin{eqnarray}
\nabla _ {\mu}\, z^{\a} &=&  0 ;\label{g1}\\
G^{-i}_{\mu \nu} \gamma^{\mu
\nu}=0 ;&&\label{g2} \\
W^{\a 11} =  0 \,; &\quad & W^{i21}=0 \label{g3}\\
\G^{\a}_{\ ij} = g^{\a\bar\b}\partial_i g_{j\bar\b} = 0 \, &\quad
& \G^{i}_{\ \a
j} g^{i\bar k}\partial_j g_{\a\bar k}=0\label{g4}\\
C_{\bar{\jmath}\bar{k}\bar{\a}}=0 \, &\quad
&C_{\bar{\b}\bar{\g}\bar{\delta}} =0 \label{g5}
\end{eqnarray}
We analyze here the constraints coming from the 3-fermions terms.
Equation (\ref{g2}) is automatically satisfied, on the reduced
theory, because of equations (\ref{spec1}) - (\ref{spec3}) and
(\ref{spec4}). Equations (\ref{g4}) are also true on the
sub-manifold since we have found in (\ref{mistametr}) that the
mixed components of the metric are zero. Finally, equations
(\ref{g5}) give constraints on the Special-Kahler manifold to be
satisfied on the reduced sub-manifold $\cM_R$. In particular, the
first one $C_{\bar{\jmath}\bar{k}\bar{\a}}=0 |_{\cM_R}$ coincides
with the already found condition (\ref{spec4}), while
$C_{\bar{\b}\bar{\g}\bar{\delta}}=0|_{\cM_R}$ is a further
constraint, due to supersymmetry, to be satisfied. We note that in
particular it implies the following constraint on the curvature of
the special manifold:
\begin{equation}
R^{\bar \imath}_{\phantom{i}\bar \a \b \bar\g}=0
\end{equation}
\vskip 5mm
 If we now define:
\begin{equation}
\chi^i \equiv \l^{i1}\,\,,\, \quad \l^\L \equiv -2 f^\L_\a
\l^\a_{\bullet}
\end{equation}
and apply the special geometry relation:
\begin{equation}
C_{\mathcal IJK} =f^{\bf \L}_{\mathcal I}\partial_{\mathcal J}
\cN_{{\bf \L}{\bf \S}}f^{\bf \S}_{\mathcal K},
\end{equation}
which gives:
\begin{equation}
C_{i\a\b} =f^{\L}_{\a}\partial_{i} \cN_{{ \L}{\S}}f^{ \S}_{\b},
\end{equation}
we can rewrite equations (\ref{gauchirtrasf}) and
(\ref{gaugautrasf}) as:
\begin{eqnarray}
\delta \,\chi^{i}&=&
 {\rm i}\,\nabla _ {\mu}\, z^{i}
\gamma^{\mu} \epsilon^\bullet \,+\, gW^{i11}\epsilon _\bullet
+\nonumber\\
&+& \frac{1}4\left (\partial_{j} \cK \bar\chi^{j}\epsilon_\bullet
- h.c. \right)\chi^{i} - \left (\omega_{1 |s}^{\phantom{A
|u}1}\bar\zeta^s \epsilon_\bullet +\omega_{1 |\bar s}^{\phantom{A
|u}1}\bar\zeta^{\bar s} \epsilon^\bullet \right)\chi^{i}
-\G^{i}_{\ jk }\bar\chi^{k}\epsilon_\bullet \chi^{j}
+\nonumber\\
&-&{\rm i}\chi^{i}\psi_{\mu \bullet}\g^\mu\epsilon^\bullet +
\frac{\rm i}8
g^{i\bar{\jmath}}\,\partial_{\bar{\jmath}}\cN_{\L\S}\bar \l^{\L
\bullet}
\l^{\S\bullet}\epsilon_\bullet \label{gauchirtrasf2}\\
\delta \,\lambda^{\L}_\bullet&=&
 F^{-\L}_{\mu \nu} \gamma^{\mu
\nu} \epsilon _\bullet \,-2\, gf^\L_\a W^{\a 21}\epsilon _\bullet
+\nonumber\\
&+& \frac{1}4\left (\partial_{j} \cK \bar\chi^{j}\epsilon_\bullet
- h.c. \right)\lambda^{\L}_\bullet-\left (\omega_{2
|s}^{\phantom{A |u}2}\bar\zeta^s \epsilon_\bullet +\omega_{2
|\bar s}^{\phantom{A |u}2}\bar\zeta^{\bar s}\epsilon^\bullet
\right)\lambda^{\L}_\bullet -\G^{\a }_{\ \b i}\bar\chi^{i
}\epsilon_\bullet \lambda^{\L}_\bullet
+\nonumber\\
&+& \frac{\rm i}4\left( {\rm
Im}\cN\right)^{-1\L\S}\left(\partial_{\bar{\jmath}}
\cN_{\S\G}\bar\chi^{\bar\jmath}\l^{\G\bullet}+\partial_{i}
\bar\cN_{\S\G}\bar\chi^{i}\l^{\G}_\bullet\right) \epsilon_\bullet
. \label{gaugautrasf2}
\end{eqnarray}
which have the form of the $N=1$ transformation laws for chiral-
and vector-multiplets fermions respectively.
\\
We still have to identify the precise form of the bosonic
quantities $W^{i11}$, $f^\L_\a W^{\a 21}$ and $\cN_{\L\S}$ in the
reduced theory. This has been done in \cite{lungo}. We just quote
the main results here. \\
 For example, in order to retrieve the $D$-term of the $N=1$ gaugino transformation,
 we have to identify:
\begin{equation}
\label{dterm} -2 g f^\L_\a W^{\a 21} \equiv {\rm i} D^\L  = {\rm
i} ({\rm Im }\cN^{-1})^{\L\S} \left(P^0_\S + P^3_\S\right).
\end{equation}
\par
Moreover, in order to show that equation (\ref{gaugautrasf2}) is
the correct $N=1$ transformation law of the gauginos we have
still to prove that $\cN_{\L\S}$ is an antiholomorphic function
of the scalar fields $z^i$, since the corresponding object of the
$N=2$ special geometry $\cN_{{\bf \L}{\bf \S}}$ is not. For this
purpose we observe that in $N=2$ special geometry the following
identity holds (at least when a $N=2$ prepotential function
exists)\cite{cdf}:
\begin{equation}\label{inversusf1}
{\mathcal N}_{{\bf \L}{\bf \S}}\,=\,{\bar F}_{{\bf \L}{\bf
\S}}-2{\rm i}\bar T_{{\bf \L}} \bar T_{{\bf \S}} (L^{{\bf \G}
}{\rm Im}F_{{\bf \G} {\bf \Delta}}L^{{\bf \Delta}})
\end{equation}
where the matrix $F_{{\bf \L}{\bf \S}}$ is holomorphic and $T_\L$ is the so-called
projector on the graviphoton \cite{lungo}, \cite{cdf}.   \\
If we now reduce the indices ${\bf \L}{\bf \S}$ we find:
\begin{equation}\label{holn}
{\mathcal N}_{\L\S}\,=\,{\bar F}_{\L\S}-2{\rm i}\bar T_\L \bar
T_\S (L^{Z }{\rm Im}F_{Z W}L^{W})\equiv {\bar F}_{\L\S}
\end{equation}
since, as shown in \cite{lungo}, $T_\L=0$ is precisely the bosonic
constraint derived from (\ref{tmunu}). Therefore ${\mathcal
N}_{\L\S}$ is antiholomorphic and the $D$-term (\ref{dterm})
becomes:
\begin{equation}\label{dtrm1}
  D^\L  \equiv 2{\rm i} f^\L_\a W^{\a 21} = - 2({\rm Im }f^{-1}(z^i))^{\L\S}
\left(P^0_\S + P^3_\S\right).
\end{equation}
where we have defined
\begin{equation}\label{norma}
F_{\L\S}(z^i)=\half f_{\L\S}(z^i)
\end{equation}
in order to match the normalization of the holomorphic kinetic
matrix of the $N=1$ theory appearing in equation
(\ref{dlambda}).\\
The $N=1$ transformation law of the gravitino with these
notations takes the final form:
\begin{eqnarray}
\label{grave} \delta\,\psi _{\bullet \mu} &=& {\mathcal D}_\mu
\epsilon_\bullet  -\frac{\rm i}2 \left(\hat{\cal Q}_\mu  -2{\rm i}
\hat{\omega}_{\mu | 1}^{\phantom{\mu |A}1}
\right)\epsilon_\bullet \,
 +  {\rm i} \, g \,S_{11}\eta _{\mu \nu}
 \gamma^{\nu}\epsilon^\bullet
+ \nonumber\\
&+&\frac {\rm i}4\left [{\rm
Im}f_{\L\S}\bar\l^{\L\bullet}\g_\mu\l^\S_\bullet+\half\g_{\mu\nu}
\left( \bar f_{\L\S}\bar\l^{\L\bullet}\g^\nu\l^\S_\bullet
+g_{i\bar\jmath}\bar\chi^{\bar\jmath}\g^\nu\chi^i + g_{s\bar  s
}\bar\zeta^{\bar s}\g^\nu\zeta^s
 \right)\right]\epsilon_\bullet  +\nonumber\\
&-&\left (\omega_{1 |s}^{\phantom{A |u}1}\bar\zeta^s
\epsilon_\bullet +\omega_{1 |\bar s}^{\phantom{A
|u}1}\bar\zeta^{\bar s}\epsilon^\bullet \right)\psi_{\mu \bullet
}- \frac{1}4\left (\partial_{j} \cK \bar\chi^{j}\epsilon_\bullet
- h.c. \right)\psi_{\mu \bullet }
\end{eqnarray}

\section{The gauging}
As it was stressed in the introduction, the implications of the
$N=2 \to N=1$ reduction on the gauging of the $N=2$ theory cannot
be obtained by looking only at the fermionic sectors, since the
fermionc shifts are built up in terms of bosonic fields only. The
analysis of the gauging has been thoroughly given in \cite{lungo}.
To make the paper self-contained, we just summarize here the
conclusions, and in particular:
\begin{itemize}
\item{
The $D$-term of the $N=1$-reduced gaugino $\lambda^\L = -2 f^\L_i
\l^{i2} $ is:
\begin{equation}
D^\L = W^{i21} =-2g_{(\L)}({\rm Im}f)^{-1 \L\S} \left(P^3_\S(w^s)
+P^0_\S (z^i)\right)\label{d2}
\end{equation}
}
\item{ The $N=1$-reduced superpotential, that is the gravitino mass, is:
\begin{equation}
L(z,w)=\frac{\rm i}{2}g_{(X)}L^{X} \left( P^1_{X} - {\rm i}P^2_{X}
\right) \label{grashift}
\end{equation}
and is  a holomorphic function of its coordinates $z^i$ and
$w^s$.}
 \item{The fermion shifts of the $N=1$ chiral spinors $\chi^i =\l^{i1}$
coming from the $N=2$  gaugini are:
\begin{equation}
g W^{i11} \equiv N^i =2  g^{i\bar\jmath}\nabla_{\bar\jmath}\bar L
\label{chir1shift}
\end{equation}
}
\item{The fermion shifts of the $N=1$ chiral spinors $\zeta^s$
coming from $N=2$  hypermultiplets are:
\begin{equation}
N^s =-4 g_{(X)}k^t_{X} \bar L^{X} {\mathcal U}^{1\dot I}_t
{\mathcal U}_{2\dot I}^s = 2 g^{s\bar s}\nabla_{\bar s}\bar L
.\label{chir2shift}
\end{equation}
}
\item{Only some components of the special--K\"ahler and quaternionic
prepotentials and of the corresponding Killing vectors remain
different from zero after the reduction, in particular we have:
\begin{eqnarray}
&& P^0_{X} = 0 \, , \label{ks1}\\
&& k^i_{X} =0 \, , \quad k^\a_\L =0\label{ks2}\\
 && k^i_\L = {\rm i}g^{i\bar\jmath} \nabla_{\bar\jmath} P^0_\L \neq 0 .
 \label{ks3}
\end{eqnarray}
and
\begin{eqnarray}
&& P^3_{X} = 0 \, , \quad P^{\rm i}_{\L} =0 \, , \quad ({\rm i} =1,2)\label{kq1}\\
&& k^s_{X} =0 \, , \quad k^t_\L =0\label{kq2}\\
 && k^s_\L = {\rm i}g^{s\bar s} \nabla_{\bar s} P^3_\L \neq 0 .\label{kq3}
\end{eqnarray}
We note that, by using the quaternionic relation:
\begin{equation}
n_H P^x_{\bf \L} =-\half \Omega^x_{uv}\nabla^u k^v_{\bf \L}
\end{equation}
from equation (\ref{kq1}) it follows:
\begin{eqnarray}
n_H P^{\rm i}_{ \L}& =& 0=-\half \Omega^{\rm i}_{st}\left(\nabla^s
k^t_{\L}
- \nabla^t k^s_{\L}\right) \\
n_H P^{3}_{X}& =& 0= -\half \Omega^{3}_{s\bar s}\nabla^s k^{\bar
s}_{X}+ \Omega^{3}_{t\bar t}\nabla^t k^{\bar t}_{X}
\end{eqnarray}
satisfied for:
\begin{eqnarray}
\nabla^s k^t_{\L}=0 \, ;&& \nabla^s k^{\bar s}_{X}=0 \label{kq5}\\
\nabla^t k^s_{\L} =0 \, ;&&\nabla^t k^{\bar t}_{X}=0.\label{kq6}
\end{eqnarray}
 Equations (\ref{kq5}) follow from (\ref{kq2}) for consistency
 of the reduction
 to the submanifold $\cM^{KH}$. Equations (\ref{kq6}) are instead
 further relations to be satisfied for the truncation.
One can see for instance that the above relations  do indeed hold
in the model of reference \cite{fgkp2} where the gauge group acts
linearly on the coordinates of the scalar manifolds.
 }
\end{itemize}

\section{A closer look to consistency:
 Yukawa interactions}
It is well known that, in order to have a consistent reduction,
the solutions of the equations of motion of the reduced theory
must be also solutions of the mother theory. This fact in
particular implies that all terms in the lagrangian bilinear in
the fermions, containing one retained and one truncated out
fermion, must disappear in the reduction. Indeed, the
corresponding field equations obtained by varying the lagrangian
with respect to the truncated fermion would be inconsistent. Let
us check that the bosonic quantities which are coefficients of
these terms do indeed vanish in the reduction.
\par
We will confine to analyze the ``mass'' terms of the $N=2$
lagrangian, namely:
\begin{eqnarray}
{\cal L}^{N=2}_{mass}&=& g[2 S_{AB} \bar\psi^A_\mu
\g^{\mu\nu}\psi^B_\nu + {\rm i} g_{{\mathcal I}\bar {\mathcal J}}
W^{{\mathcal I} AB} \bar\lambda^{\bar{\mathcal  J}}_A\gamma_\mu
\psi_B^\mu+
 2{\rm i} N^A_\alpha\bar\zeta^\alpha\gamma_\mu \psi_A^\mu \nonumber \\
&+& {\cal M}^{\alpha\beta}{\bar \zeta}_\alpha \zeta_\beta +{\cal
M}^{\alpha}_{\phantom{\alpha}{\mathcal I}B} {\bar\zeta}_\alpha
\lambda^{{\mathcal I}B} + {\cal M}_{{\mathcal I}{\mathcal J}\,AB}
{\bar \lambda}^{{\mathcal I}A} \lambda^{{\mathcal J} B} +
\mbox{h.c.}]
\end{eqnarray}
where, besides the matrices $S_{AB}$, $W^{{\mathcal I} AB}$,
$N^A_\alpha$ defined in (\ref{trapsi2}) - (\ref{traqua2}), there
appear the mass matrices (see \cite{df}):
\begin{eqnarray}\label{mass2}
{\cal M}^{\alpha\beta}  &=&- \, {\cal U}^{\alpha A}_u \, {\cal
U}^{\beta B}_v \, \varepsilon_{AB}
\, \nabla^{[u}   k^{v]}_{\bf \Lambda}  \, L^{\bf \Lambda} =
-\half {\mathcal U}^{A\a |u} \nabla_u N^\b_A \\
{\cal M}^{\alpha }_{\phantom{\alpha} {\mathcal I}B} &=& -4 \,
{\cal U}^{\alpha}_{B  u} \, k^u_{\bf\Lambda} \,
 f^{\bf\Lambda}_{\mathcal I} \\
{\cal M}_{AB \,\, {\mathcal I}{\mathcal K}} &=&  \,  \epsilon_{AB}
\, g_{\bar{\mathcal L} [{\mathcal I}}
 f_{{\mathcal K}]}^\Lambda  k^{\bar {\mathcal L}}_ \Lambda
 \,+ \frac {1}{2}
{\rm {i}}  P_ {{\bf\Lambda} AB} \, \nabla_{\mathcal I}
f^{\bf\Lambda} _{\mathcal K} \label{pesamatrice}
\end{eqnarray}
 The gravitino mass term $S_{12}\bar\psi^1_\mu
\g^{\mu\nu}\psi^2_\nu$ is automatically zero because of
(\ref{s21}).
\\
The term $g_{{\mathcal I}\bar {\mathcal J}} W^{{\mathcal I} AB}
\bar\lambda^{\bar{\mathcal  J}}_A\gamma_\mu \psi_B^\mu$ contains
four potentially dangerous contributions, namely:
\begin{eqnarray}
&&g_{i\bar\jmath} W^{i12} \bar\lambda^{
\bar \jmath}_1\gamma_\mu \psi_2^\mu \\
&&g_{i\bar\jmath} W^{i12} \bar\lambda^{
\bar \jmath}_2\gamma_\mu \psi_1^\mu \\
&&g_{\a\bar \b} W^{\a 11} \bar\lambda^{ \bar \b}_1\gamma_\mu
\psi_1^\mu\\
&&g_{\a\bar \b} W^{\a 22} \bar\lambda^{ \bar \b}_2\gamma_\mu
\psi_2^\mu
\end{eqnarray}
Looking at the expression (\ref{tralam}) of $W^{{\mathcal I} AB}$
we see that $W^{i12}$ is zero on the reduced theory, taking into
account the constraints: $P^{12}_{\bf \L} \equiv P^3_\L$, $f^\L_i
=0$, $L^{\bf \L} \equiv L^X$, $k^i_X=0$; furthermore, $W^{\a 11}$
and $W^{\a 22}$ are both zero due to the constraints: $P^{11}_{\bf
\L} =( P^{22}_{\bf \L})^*\equiv P^1_X-{\rm i}P^2_X$, $f^X_\a =0$.
\\
Then, we need the terms $N^1_{\dot I} \bar\zeta^{\dot I}
\gamma_\mu \psi_1^\mu $ and $N^2_I\bar\zeta^I\gamma_\mu \psi_2^\mu
$ to be zero. And indeed, $N^1_{\dot I} = (N^2_I)^\star
=0|_{\cM_R}$ for consistency of the truncation of half
hypermultiplets (see equation (\ref{hyp3})).\\
Furthermore, the term ${\cal M}^{I\dot J}{\bar \zeta}_I
\zeta_{\dot J}$ must be zero, and this is satisfied if:
\begin{equation}
{\cal M}^{I\dot J} = \, {\cal U}^{I 1}_{\bar s} \, {\cal U}^{\dot
J 2}_s \, \nabla^{[\bar s } k^{s]}_{ X}  \, L^{X} -\, {\cal U}^{I
2}_{\bar t} \, {\cal U}^{\dot J 1}_t \, \nabla^{[\bar t } k^{t]}_{
X}  \, L^{X}=0.\label{mij}
\end{equation}
Both terms are indeed zero as a consequence of (\ref{spec1}) and
(\ref{kq1}),(\ref{kq2}),(\ref{kq6}). \\
From the mixing term ${\cal
M}^{\alpha}_{\phantom{\alpha}{\mathcal I}B} {\bar\zeta}_\alpha
\lambda^{{\mathcal I}B} $ we get the potentially inconsistent
contributions:
\begin{eqnarray}
&&{\cal M}^{I}_{\phantom{\alpha}{\a 1}}
{\bar\zeta}_I\lambda^{\a 1};\label{mix1}\\
&&{\cal M}^{I}_{\phantom{\alpha}i2} {\bar\zeta}_I
\lambda^{i2};\label{mix2}\\&&{\cal M}^{\dot I
}_{\phantom{\alpha}i1} {\bar\zeta}_{\dot I}
\lambda^{i1};\label{mix3}\\&&{\cal M}^{\dot I
}_{\phantom{\alpha}\a 2} {\bar\zeta}_{\dot I} \lambda^{\a 2
};\label{mix4}
\end{eqnarray}
All these quantities are indeed zero on the reduced theory, as can
be ascertained by using in (\ref{mass2}) the relations
(\ref{surv}),  (\ref{surv2}), (\ref{spec1}), (\ref{spec2}) and
(\ref{kq2}).
\\
Finally, the gaugino mass term ${\cal M}_{{\mathcal I}{\mathcal
J}\,AB} {\bar \lambda}^{{\mathcal I}A} \lambda^{{\mathcal J} B} $
contains the four contributions:
\begin{eqnarray}
&&{\cal M}_{ij\,12} {\bar
\lambda}^{i1} \lambda^{j2} \\
&&{\cal M}_{\a\b \,12} {\bar
\lambda}^{\a 1} \lambda^{\b 2}\\
&&{\cal M}_{i\a \,11} {\bar \lambda}^{i1} \lambda^{\a 1} \\
&&{\cal M}_{i\a \,22} {\bar \lambda}^{i 2} \lambda^{\a 2}
\end{eqnarray}
which have to be zero on the truncated theory. We find:
\begin{eqnarray}
{\cal M}_{ij\,12} &=& g_{\bar k [i}f^X_{j]}k^{\bar k}_X -
\frac{\rm i}2P_{\L 12}\nabla_i f^\L_j \label{m1}\\
{\cal M}_{\a\b \,12} &=& g_{\bar \g [\a}f^\L_{\b]}k^{\bar \g}_\L -
\frac{\rm i}2P_{\L 12}\nabla_\a f^\L_\b\label{m2}\\
{\cal M}_{i\a \,11} &=& -
\frac{\rm i}4 P_{X 11}\nabla_\a f^X_i \label{m3} \\
{\cal M}_{i\a \,22}&=& - \frac{\rm i}4 P_{X 22}\nabla_\a f^X_i
\label{m4}
\end{eqnarray}
Equations (\ref{m1}) -- (\ref{m4}) are all satisfied as a
consequence of (\ref{spec1}), (\ref{spec2}), (\ref{spec4}),
(\ref{g5}) and (\ref{ks2}), (\ref{kq1}).
\par
\vskip 5mm Finally, coming to the reduction of the scalar
potential of the $N=2$ theory down to $N=1$, we have that the
$N=2$ scalar potential, given by:
\begin{equation}\label{pot2}
  \cV^{N=2}
 =\left( g_{{\mathcal I}\bar{\mathcal J}}k^{{\mathcal I}}_{{\bf \L }} k^{\bar{\mathcal J}}_{{\bf \S }} +
 4h_{uv}k^u_{{\bf \L }}
 k^v_{{\bf \S }}\right)\bar L^{{\bf \L }} L^{{\bf \S }} + \left(- \frac {1} {2} ({\rm
 Im}{\mathcal N^{-1})}^{{\bf \L }{\bf \S }} -\bar L^{{\bf \L }} L^{{\bf \S }}\right)
 P^x_{{\bf \L }}
 P^x_{{\bf \S }}
-3P^x_{{\bf \L }} P^x_{{\bf \S }} {\bar L}^{{\bf \L }} L^{{\bf \S
}}
\end{equation}
reduces to the standard form for the $N=1$ scalar potential,
written in terms of the covariantly holomorphic superpotential
$L$ as:
\begin{equation} \label{21pot} {\cal V}^{N=2 \to N=1} = 4
\left[ -3 L \bar L + g^{i\bar\jmath} \nabla_i L
\nabla_{\bar\jmath} \bar L + g^{s\bar s} \nabla_s L \nabla_{\bar
s} \bar L + \frac{1}{16} {\rm Im} f_{\L\S} D^\L D^\S \right]
\end{equation}
The explicit proof is given in \cite{lungo}.

\end{document}